# Plasma flow generation and particle acceleration from expanding magnetic bubbles

Yang Zhang [1,2,*] Brandon K. Russell [1] Geoffrey Pomraning [1] Lan Gao [3] Xiaocan Li [4] Adam Stanier [4] William Daughton [4] Chuanfei Dong [5,6] Liang Wang [5] Peiyun Shi [3] Kian Orr [7] and Hantao Ji [1,3]

[1] *Department of Astrophysical Sciences, Princeton University, Princeton, New Jersey 08544, USA*
[2] *University Corporation for Atmospheric Research, Boulder, Colorado 80301, USA*
[3] *Princeton Plasma Physics Laboratory, Princeton University, Princeton, New Jersey 08543, USA*
[4] *Los Alamos National Laboratory, Los Alamos, New Mexico 87545, USA*
[5] *Center for Space Physics and Department of Astronomy, Boston University, Boston, Massachusetts 02215, USA*
[6] *School of Natural Sciences, Institute for Advanced Study, Princeton, New Jersey 08540, USA*
[7] *Department of Mechanical and Aerospace Engineering, Princeton University, Princeton, New Jersey 08544, USA*

Impulsive plasma dynamics in the laboratory are often driven by rising electric currents, yet their quantitative plasma response has not been well established. By means of fully kinetic particle-in-cell simulations and laser-driven capacitor-coil experiments, we show that a rising current expels plasma, forming an expanding magnetic bubble and accelerating particles. The expansion front velocity scales with the Alfvén speed determined by the magnetic field at its inner edge and the plasma density at its outer edge. This mechanism establishes impulsive current drive as a fundamental way that generates plasma flows and accelerates particles in laboratory plasmas, with potential relevance to astrophysics.

## I. INTRODUCTION

Electric currents are direct and versatile control parameter in plasma physics laboratories. By driving currents with pulsed-power machines, high-intensity lasers, or external coils, experiments can generate strong magnetic fields [1,2], drive plasma pinching and expansion [3–6], launch flows, jets, and shocks [7–11], and initiate magnetic reconnection [12–18] under controlled and repeatable conditions. While these experiments have revealed many fundamental plasma processes, the plasma's direct response to a rising current has received less systematic attention. A time-varying current is not simply a tool for shaping magnetic geometry—It is a fundamental driver that exerts forces, redistributes plasma, and accelerates particles. Yet, the quantitative response—including flow generation and particle energization—remains underexplored. A quantitative understanding of this response is crucial for revealing the fundamental role of impulsive current drive in shaping plasma evolution across laboratory experiments and may offer insights relevant to astrophysical environments.

In this article, we present a combined simulation-theory-experiment demonstration of the fundamental plasma response to an increasing current, establishing both a scaling law and experimental validation with direct diagnostic signatures. Fully kinetic particle-in-cell simulations reveal that a linearly rising current spontaneously induces a reverse current of equal magnitude and opposite direction in the surrounding plasma. The mutual repulsion between these currents expels plasma, driving the formation of an expanding magnetic bubble. The bubble front velocity follows a scaling, $u = 0.28(\frac{\mu_0 I_0^2}{\rho_0 \tau^2})^{1/4}$, where $I(t) = I_0 t/\tau$ is the imposed current profile, $\mu_0$ is the vacuum permeability, and $\rho_0$ is the background plasma mass density. This expansion also energizes particles: The ion spectrum develops a tail and two distinct peaks—one at the bulk expansion velocity and another near twice this value—while the electron spectrum exhibits a strong tail driven by the inductive electric field sustaining the reverse current. These features are confirmed experimentally in laser-driven capacitor-coil targets, where proton radiography captures magnetic structures consistent with reverse-current formation and expansion velocities close to the predicted scaling, and direct electron spectra measurements show enhanced suprathermal generation. The consistency between first-principles simulations and experimental diagnostics highlights a general plasma response mechanism to time-varying currents, with broad implications for impulsive magnetic activity in both laboratory and astrophysical systems.

## II. SIMULATION RESULTS

We performed fully kinetic simulations using VPIC [19], with parameters motivated by laser-driven capacitor-coil experiments [2,20–22]. The two-dimensional simulation domain spans 1.6 mm × 1.6 mm, bounded by conducting walls. In the center, a wire with a radius $r_w = 0.025$ mm serves as an inner boundary carrying a time-dependent current $I = I_0 t/\tau$ along the negative $z$ direction, where $I_0 = 100$ kA is the peak current and $\tau = 0.5$ ns is the rise time. The background plasma is uniform with electron density $n_e = 10^{23}$ m$^{-3}$. The ion is copper with mass $m_i = 63.55m_p$ and charge state $Z = 18$. Both electrons and ions are initialized at $T_e = T_i = 400$ eV. The simulation domain is divided into $1696 \times 1704$ cells with 200 particles per cell per species. The corresponding characteristic plasma parameters are the electron skin depth $d_e = 0.016$ mm, the ion skin depth $d_i = 1.4$ mm, and the electron plasma frequency $\omega_{pe} = 1.8 \times 10^{13}$ s$^{-1}$.

*Contact author: yz0172@princeton.edu

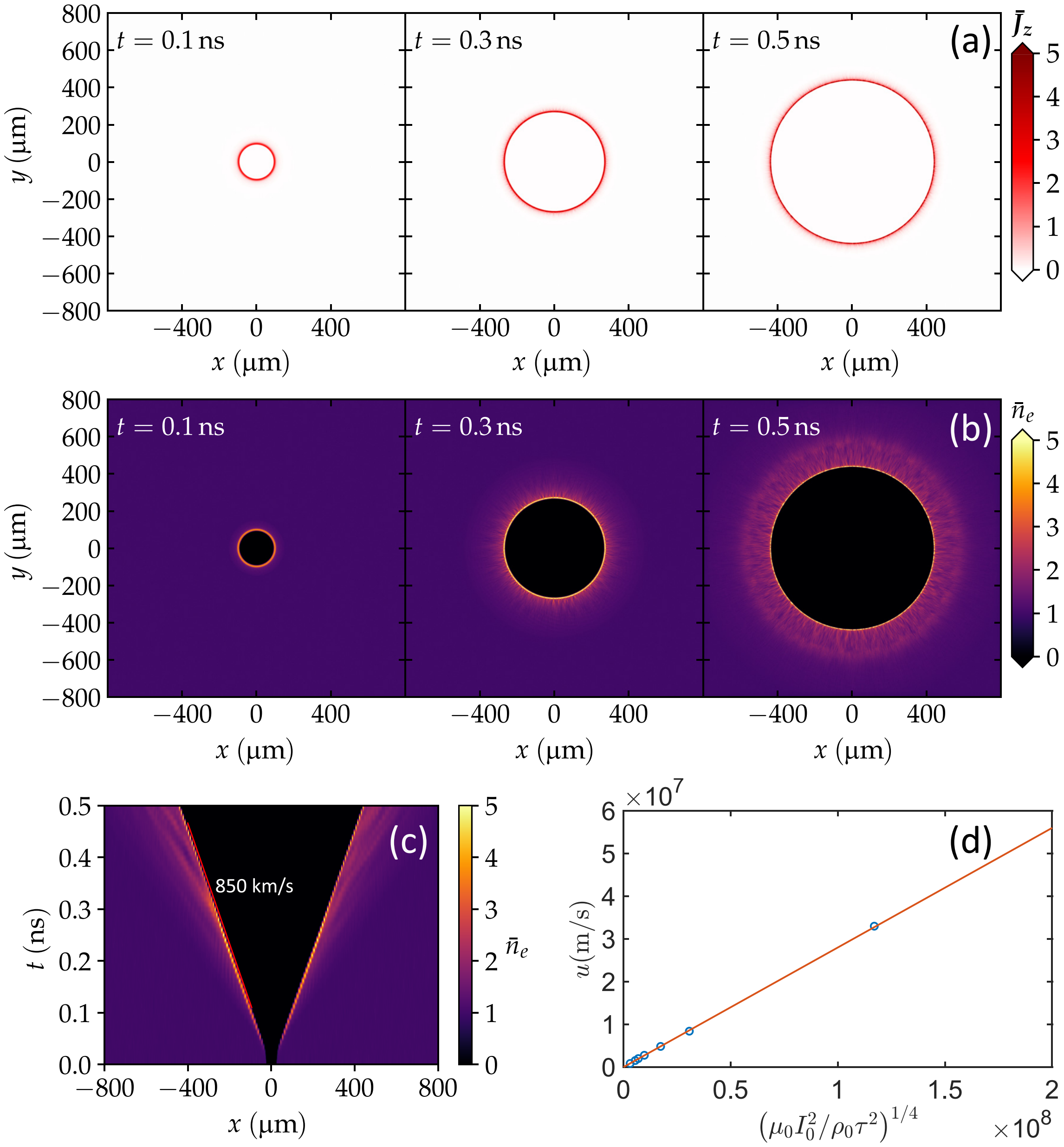

FIG. 1. Plasma response to a rising wire current. (a) Normalized axial current density at $t = 0.1$, 0.3, and 0.5 ns corresponding to $w_{pe}t =$ 1782, 5346, and 8910 (central wire omitted). A thin reverse-current layer forms and expands outward. (b) Normalized electron density at $t = 0.1$, 0.3, and 0.5 ns, showing plasma expulsion from the coil and the formation of a void region. (c) Contour of density along $y = 0$ in the $x$–$t$ plane, highlighting the accelerating plasma front that saturates at 850 km/s. (d) Front velocity from simulations with varied current, rise time, and background density. Blue points denote simulation results, and the orange line shows the fitted scaling $u = 0.28(\mu_0 I_0^2/\rho_0\tau^2)^{1/4}$.

Figure 1 presents the plasma response to a rising wire current, where a reverse current is spontaneously generated in the surrounding plasma and exerts a current-current repulsion: the formation of an expanding plasma front that leaves a near-vacuum void behind it. Figure 1(a) shows the normalized axial current density $\bar{J}_z = J_z/J_0$ distribution at $t = 0.1$, 0.3, and 0.5 ns, where $J_0 = 1$ A/μm$^2$. The central wire current density is omitted for clarity. A reverse-current ring forms around the wire at a finite radius. This reverse current is concentrated in a thin shell that expands outward over time. We quantify the total current by integrating over the coil and the reverse-current layer. The reverse current has the same magnitude but opposite direction as the coil current, as shown in Fig. 4(a), so the net enclosed current outside the layer is zero. Defining the inner radius of the reverse-current ring as $R$ and the thickness of the ring as $\delta$, the magnetic-field distribution naturally is divided into three regions, as illustrated in Fig. 4(b). In Region I ($r_w < r < R$), it follows a $1/r$ decay corresponding to the central wire's vacuum field. In Region II ($R < r < R + \delta$), the decay steepens as the reverse current cancels the wire current. In Region III ($r > R + \delta$), the field vanishes because the enclosed net current is zero. The current within the reverse-current layer is primarily carried by electrons, as demonstrated in Fig. 4(c). The electron drift arises from $\mathbf{E} \times \mathbf{B}$ drift motion, established by the balance between the radial electric force and magnetic Lorentz force, which sustains an axial drift supporting the return current.

Figure 1(b) shows the electron density at $t = 0.1$, 0.3, and 0.5 ns. The plasma is expelled from the coil and accumulates at an outward-propagating front. This process leaves behind a density void—an extended region of near-zero density—between the central wire and the expanding accumulation front. Figure 1(c) presents a contour plot of the electron density along $y = 0$ in the $x$–$t$ plane that tracks the motion of the plasma front along $y = 0$. The front accelerates at early times and then reaches a constant velocity of 850 km/s. To quantify the dependence of this expansion on system parameters, we performed additional simulations varying the peak current, rise time, and background density. Figure 1(d) summarizes these results, showing that the measured front velocities follow the scaling relation

$$u = 0.28\left(\frac{\mu_0 I_0^2}{\rho_0 \tau^2}\right)^{1/4}, \tag{1}$$

where $\rho_0$ is the background plasma mass density. This expression demonstrates a dependence of the expansion velocity on both the current drive and the plasma mass density. A simple force-balance model (see Appendix A) yields the same scaling.

The scaling velocity is closely related to the Alfvén velocity at the moving front. The magnetic field at the inner edge of the front is

$$B_\theta = \frac{\mu_0 I(t)}{2\pi R(t)}. \tag{2}$$

Using $R(t) = ut$ and $I(t) = I_0 t/\tau$, this becomes

$$B_\theta = \frac{\mu_0 I_0}{2\pi u\tau}. \tag{3}$$

Defining the Alfvén velocity based on this field as

$$v_{A0} = \frac{B_\theta}{\sqrt{\mu_0 \rho_0}}, \tag{4}$$

and substituting the expansion velocity obtained from the simulation [Eq. (1)], we find the relation

$$u \approx 0.5 v_{A0}. \tag{5}$$

Thus, the expansion velocity equals one-half of the local Alfvén velocity defined by the magnetic field at the front's inner edge and the plasma density at its outer edge. This proportionality is physically consistent, as the expansion is magnetically driven and governed by the balance of magnetic pressure and inertia.

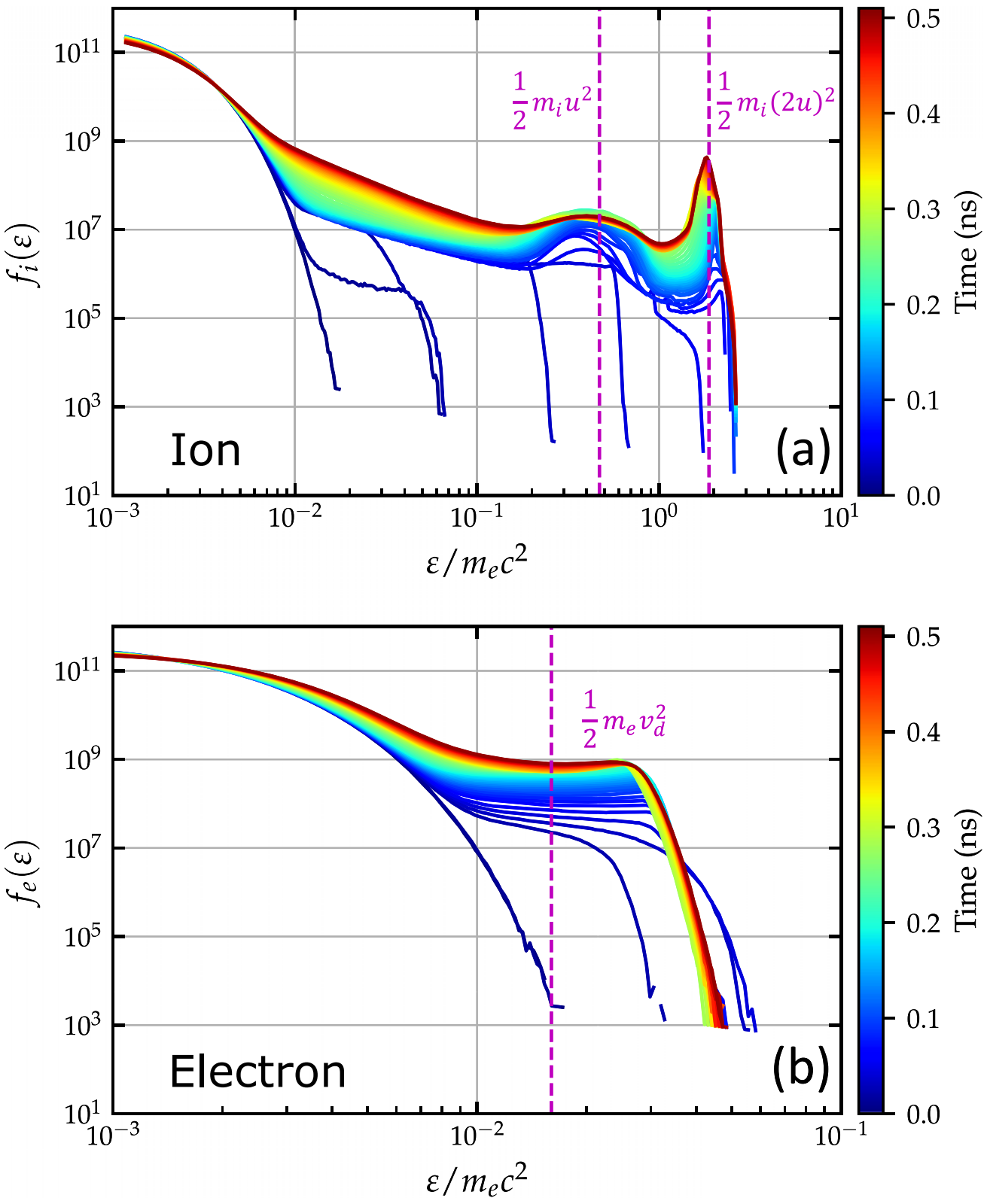

FIG. 2. Time-evolving particle energy spectra. (a) Ion distribution: The spectrum exhibits two characteristic peaks, one at the bulk expansion energy ($\frac{1}{2}m_iu^2$) and another near twice the expansion velocity ($\frac{1}{2}m_i(2u)^2$). The peaks correspond to expanding and reflected ion populations. (b) Electron distribution: The bulk expansion corresponds to only $4 \times 10^{-6}\, m_ec^2$ ($\sim$2 eV) and is not visible here. Instead, a pronounced tail develops from current-carrying electrons accelerated by the inductive electric field, which sustains the imposed reverse current.

Beyond the bulk expansion, the simulations reveal distinct energization of ions and electrons as shown in Fig. 2. The ion spectrum in panel (a) exhibits two characteristic peaks. The first corresponds to the bulk expansion energy ($\frac{1}{2}m_iu^2$), produced as ions are swept outward with the expanding front. The second peak arises near twice this expansion velocity value ($\frac{1}{2}m_i(2u)^2$), originating from ions that are reflected by the moving front: In the front frame, an incoming ion at $-u$ is reversed to $+u$, which transforms back to the laboratory frame yielding a final velocity of approximately $2u$, as illustrated in Fig. 5. This suprathermal population could potentially provide a seed for collisionless shock injection, a key step in diffusive shock acceleration [23], and may also supply energetic ions for turbulence and reconnection [24]. In contrast, the electron spectrum shows negligible bulk expansion energy ($\sim 4 \times 10^{-6} m_ec^2$) but a pronounced tail driven by the inductive electric field from the current increase, which sustains the reverse current. At $t = 0.5$ ns, the peak current density is $J_{z,\mathrm{peak}} = 4.3$ A/μm$^2$ and the peak density is $n_{e,\mathrm{peak}} = 5.0 \times 10^{23}$ m$^{-3}$, corresponding to an average electron drift velocity $v_d = J_{z,\mathrm{peak}}/(n_{e,\mathrm{peak}} e) = 0.18c$. The associated drift energy, $\frac{1}{2}m_e v_d^2 = 0.017 m_e c^2$, coincides with the tail. These results demonstrate, for the first time, that impulsive current injection alone can self-consistently generate suprathermal populations in both ions and electrons, establishing a possible link between current dynamics and particle acceleration in laboratory and astrophysical plasmas.

## III. EXPERIMENTAL VALIDATION

The reverse-current structure, expansion flow, and electron acceleration are experimentally confirmed using a laser-driven capacitor-coil setup. A capacitor-coil target consists of two parallel plates connected by a conductive wire or coil [1,2,21,25,26], with the target and experimental configuration shown in Fig. 3(a). When an intense laser irradiates the back plate, the laser-solid interaction generates electrons that are subsequently collected by the front plate. This process establishes a voltage difference between the plates, driving a current through the coil and producing strong magnetic fields. The magnetic-field structure and coil current are diagnosed using proton radiography [27]. In this technique, a proton beam produced via the target normal sheath acceleration mechanism from a separate laser-irradiated copper foil is directed across the coil, opposite to the direction of the coil current. As the protons traverse the magnetic-field region, they are deflected by the Lorentz force, producing characteristic void and deflection structures in the proton radiograph.

Figure 3(d) presents the experimental radiograph at 3.1 ns using 29 MeV protons. In the image, darker regions correspond to higher proton flux. Two concentric rings are clearly visible around the coil axis (arrows), a feature that cannot be explained by the coil current alone, which would generate only a single ring. To investigate this feature, we performed proton radiography simulations using PlasmaPy [28,29]. Figure 3(e) shows the result when only a 22-kA U-shaped coil current is included [29], producing a single concentration ring. The value of 22 kA is chosen to reproduce the observed void size. In contrast, Fig. 3(f) incorporates an additional reverse-current layer located at 0.42 mm from the coil surface, modeled by setting the magnetic field to zero at points with a minimum distance from the coil exceeding 0.42 mm. With this reverse current included, an additional outer ring emerges, in good agreement with the experimental observations.

The reverse-current position is measured to be 0.42 mm away from the coil at $t = 3.1$ ns, corresponding to an average expansion velocity of 135 km/s. An electron density of $n_e \approx 3 \times 10^{24}$ m$^{-3}$ with an average copper ion charge state of $Z = 18$ has been measured by Thomson scattering in similar experiments [17,30], corresponding to a mass density of $\rho_0 \simeq 1.7 \times 10^{-2}$ kg/m$^3$. In this experiment, the peak current reaches 32 kA with a rise time of 1 ns [2]. Using these parameters in the scaling relation of Eq. (1) yields a predicted expansion velocity of 145 km/s, in good agreement with the measurement. The observed velocity is smaller than the predicted one because the current rise time is limited to 1 ns. After 1 ns, the current begins to decay, and the magnetic force exerted by the moving current layer cannot accelerate additional plasmas to the same velocity, leading to a reduced average expansion speed between 1 and 3.1 ns. A similar reverse-current-induced two-ring structure has also been observed in other laser-driven capacitor-coil experiments [31,32]. Furthermore, reverse currents have been directly measured in other current-driven plasma experiments [33,34].

Suprathermal electron generation has also been directly measured in related experiments using a five-channel Osaka University Electron Spectrometer (OU-ESM) [30]. The time-integrated spectrometer was positioned 37.5 cm from the coil center at a polar angle of 39° and covered an azimuthal range from 179° to 199° using five equally spaced detection channels. The orientation of the OU-ESM channels is shown in Figs. 3(b) and 3(c). As shown in Figs. 3(g) and 3(h), the one-coil target produces over an order of magnitude more energetic electrons in the 30–100 keV range than the no-coil target (in which the coil is removed but the back and front plates remain, with lasers incident on the back plate), confirming that current drive enhances suprathermal electron production.

## IV. DISCUSSION

These findings demonstrate that impulsive currents provide a unifying mechanism for driving plasma flows and energizing particles across both laboratory and astrophysical plasmas.

In the laboratory, this mechanism offers a new means of generating plasma flows and energetic ions in the laboratory. Such flows are central to studies of plasma flow interactions [35–38] and shock formation [8,39–43]. Unlike the conventional laser-foil approach in high-energy-density (HED) plasmas, which relies on thermal expansion, the current-driven mechanism decouples laser-target coupling from the flow region, enabling lower-density plasmas and faster streams. For example, a picosecond laser-driven capacitor-coil target can deliver peak currents of $I_0 = 150$ kA with the current rise time taken to be comparable to the laser pulse duration $\tau = 15$ ps [21]. In a hydrogen plasma with a density of $10^{21}$ m$^{-3}$, these parameters correspond to expansion velocities of the order $0.1c$, approaching a largely unexplored relativistic regime. In addition, the mechanism accelerates ions into radially propagating, narrow energy spread suprathermal populations, offering a controllable source of directed ions with potential applications in proton radiography and beam-plasma interaction studies.

The physics explored here is directly relevant to laboratory current-driven reconnection experiments [12,30,44], where two current coils with time-dependent drive currents propel background plasma toward the reconnection site. Conceptually, this corresponds to doubling the wire number in our single-wire simulation: A rising current generates two reverse-current structures whose expansion flows converge to form the reconnection current sheet. This perspective provides insight into the initiation and sustainment of current-driven reconnection in laboratory plasmas.

Beyond the laboratory, this mechanism may underlie cavity formation, plasma flows, and particle acceleration in astrophysical systems driven by impulsive currents [45]. Such currents have been observed or inferred in solar coronal loops [46,47], astrophysical jets [48], cosmic rays [49,50], and magnetars [51]. Configurations consisting of a central current channel accompanied by distributed return currents in the surrounding plasma are invoked in astrophysical models. For example, in solar flares, rapidly evolving current channels and energetic electron beams induce return currents that regulate energy transport and plasma dynamics along magnetic loops [52–54]. In cosmic-ray-modified regions, streaming particles generate net currents that are balanced by background plasma return currents, driving magnetic-field amplification through current-driven instabilities [49,55]. Similarly, current-carrying astrophysical jets and magnetic-tower outflows consist of central current channels surrounded by diffuse return-current sheaths whose evolution leads to magnetically dominated expansion and cavity formation [56,57]. The present experiments and simulations isolate this fundamental driven-current regime in a controlled setting, enabling direct investigation of the inductively driven plasma response that shares essential electrodynamic ingredients with these diverse astrophysical environments.

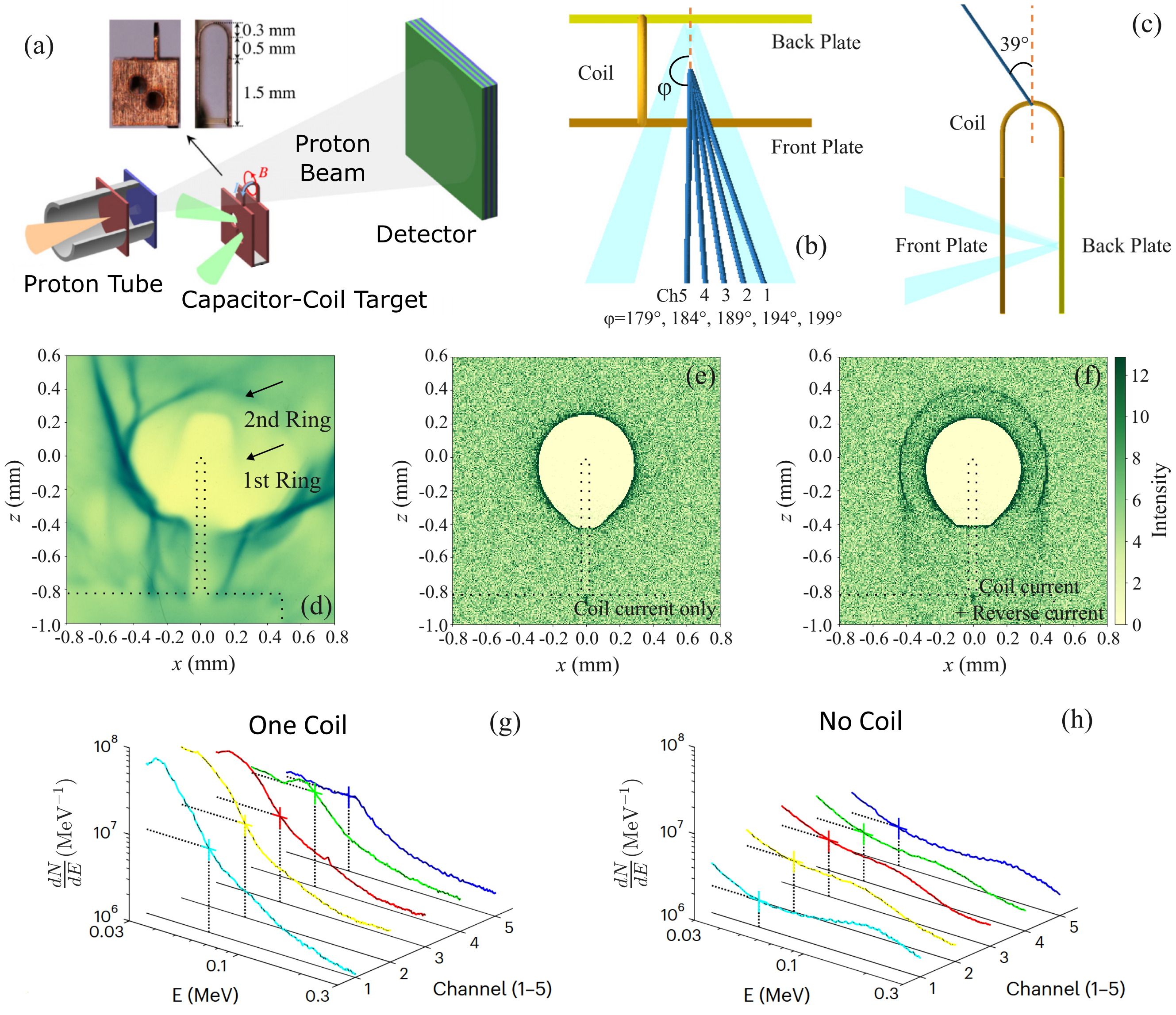

FIG. 3. Experimental validation of plasma dynamics driven by an increasing current. (a) Proton radiography setup for the laser-driven capacitor-coil experiment. Two Cu foils are connected by a U-shaped coil. Long-pulse laser irradiation generates a voltage difference between the plates, driving large currents through the coil and producing strong magnetic fields. (b), (c) Top and side views of the OU-ESM setup used for electron measurements. (b) Top-down view showing the azimuthal orientation of the OU-ESM channels, where $\varphi = 0^\circ$ is defined as the direction normal to the back plate and pointing away from the front plate. (c) Side-on view showing the relative polar orientation of the channels. A left coil is used in the particle measurement setup, while a right coil is used in the proton radiography experiment. (d) Experimental proton radiograph at $t = 3.1$ ns with 29 MeV protons, showing two concentric rings around the coil axis. (e) Simulated radiograph including only the coil current, yielding a single ring. (f) Simulated radiograph including both the coil current and a reverse-current layer, reproducing the two-ring structure observed experimentally. Electron energy spectra measured with the five-channel electron spectrometer from the one-coil target [panel (g)] and the no-coil target [panel (h)]. Crosses at 60 keV indicate representative horizontal and vertical error bars. Panels (a) and (b) are reproduced from Ref. [2], and panels (g) and (h) are reproduced from Ref. [30].

While the present analysis assumes a two-dimensional geometry, deviations from perfect axisymmetry in realistic three-dimensional configurations may modify local current and magnetic-field structures. Nevertheless, as long as the dominant dynamics are governed by the balance between magnetic forces and plasma inertia, the same velocity scaling is expected to remain qualitatively valid. Although the laboratory experiment is inherently three dimensional, the two-dimensional simulations reproduce the essential observed dynamics, indicating that the reduced dimension captures the dominant physics. Experimental evidence for analogous three-dimensional expansion layers exists in $z$-pinch studies of magnetic-tower jets, where a diffuse return-current region surrounds the central current channel [33]. In three dimensions, the reverse-current layer may be susceptible to current-driven instabilities, potentially leading to filamentation and fragmentation of the expanding plasma front. In the two-dimensional simulations, filaments are generated in the electron density, radial current density, and out-of-plane axial magnetic field. These features may arise from current driven or Weibel-type instabilities. Fully three-dimensional simulations are a natural next step to assess their potential influence on system stability and long-term evolution.

## V. SUMMARY

In summary, we have demonstrated that a rising current expels plasma to form expanding magnetic bubbles, with the expansion velocity following a scaling, $u = 0.28(\mu_0 I_0^2/\rho_0\tau^2)^{1/4}$. At the same time, the process self-consistently produces suprathermal particle populations that could potentially provide natural seeds for shocks, magnetic reconnection, and turbulence. These results establish impulsive currents as a fundamental driver of both plasma flows and particle acceleration in the laboratory, with potential relevance to astrophysical systems.

## ACKNOWLEDGMENTS

This work was supported by the NASA Living with a Star Jack Eddy Postdoctoral Fellowship Program, administered by UCAR's Cooperative Programs for the Advancement of Earth System Science (CPAESS) under Award No. 80NSSC22M0097, U.S. Department of Energy High-Energy-Density Laboratory Plasma Science program under Grant No. DE-SC0020103. C.D. was supported by the DOE Grant No. DE-SC0024639, the Alfred P. Sloan Research Fellowship, and the IBM Einstein Fellow Fund at the Institute for Advanced Study, Princeton. Y.Z. thanks Yukang Shu and Fan Guo for helpful discussions.

## DATA AVAILABILITY

The data that support the findings of this article are not publicly available. The data are available from the authors upon reasonable request.

## APPENDIX A: EXPANSION VELOCITY DERIVATION

To understand the expansion velocity scaling in Eq. (1), we develop a simple analytical model. We assume that the reverse-current layer carries a uniform current density $J_z = -I/S$, where the layer area is $S = \pi(2R\delta + \delta^2)$, and the plasma within the layer moves radially outward with a velocity $U$. The azimuthal magnetic field within the layer, for $R < r < R + \delta$, is

$$B_\theta(r) = \frac{\mu_0 I}{2\pi r}\left(1 - \frac{\pi(r^2 - R^2)}{S}\right). \tag{A1}$$

The total radial magnetic force per unit axial length is then

$$f_B = \int_R^{R+\delta} 2\pi J_z B_\theta r\, dr. \tag{A2}$$

Substituting Eq. (A1), we obtain

$$f_B = \frac{\mu_0 I^2}{3\pi}\frac{3R + 2\delta}{(2R + \delta)^2}. \tag{A3}$$

The mass per unit length of the current layer is

$$M = \rho_0\pi[(R + \delta)^2 - a^2], \tag{A4}$$

where $a$ is the coil radius. Neglecting plasma pressure compared to the magnetic force, the equation of motion is

$$\frac{d}{dt}(MU) = \frac{\mu_0 I^2}{3\pi}\frac{3R + 2\delta}{(2R + \delta)^2}. \tag{A5}$$

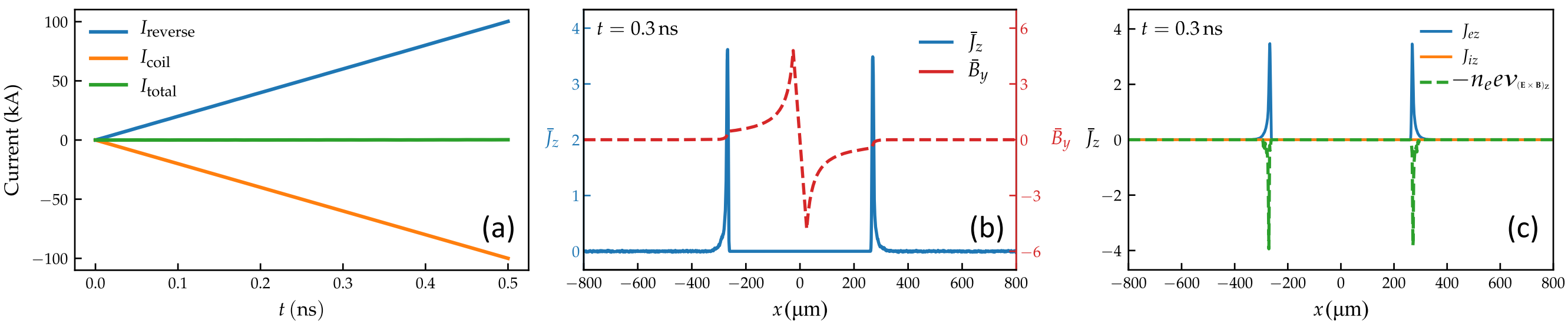

FIG. 4. Reverse current and magnetic-field structure. (a) Integrated current showing that the reverse current cancels the coil current, leaving zero net current outside. (b) Normalized axial current density and $y$-component magnetic field along $x$ at $y = 0$. (c) Normalized axial electron current density, ion current density, and electron $\mathbf{E} \times \mathbf{B}$ drift contribution, showing that the reverse current is predominantly carried by electron $\mathbf{E} \times \mathbf{B}$ motion.

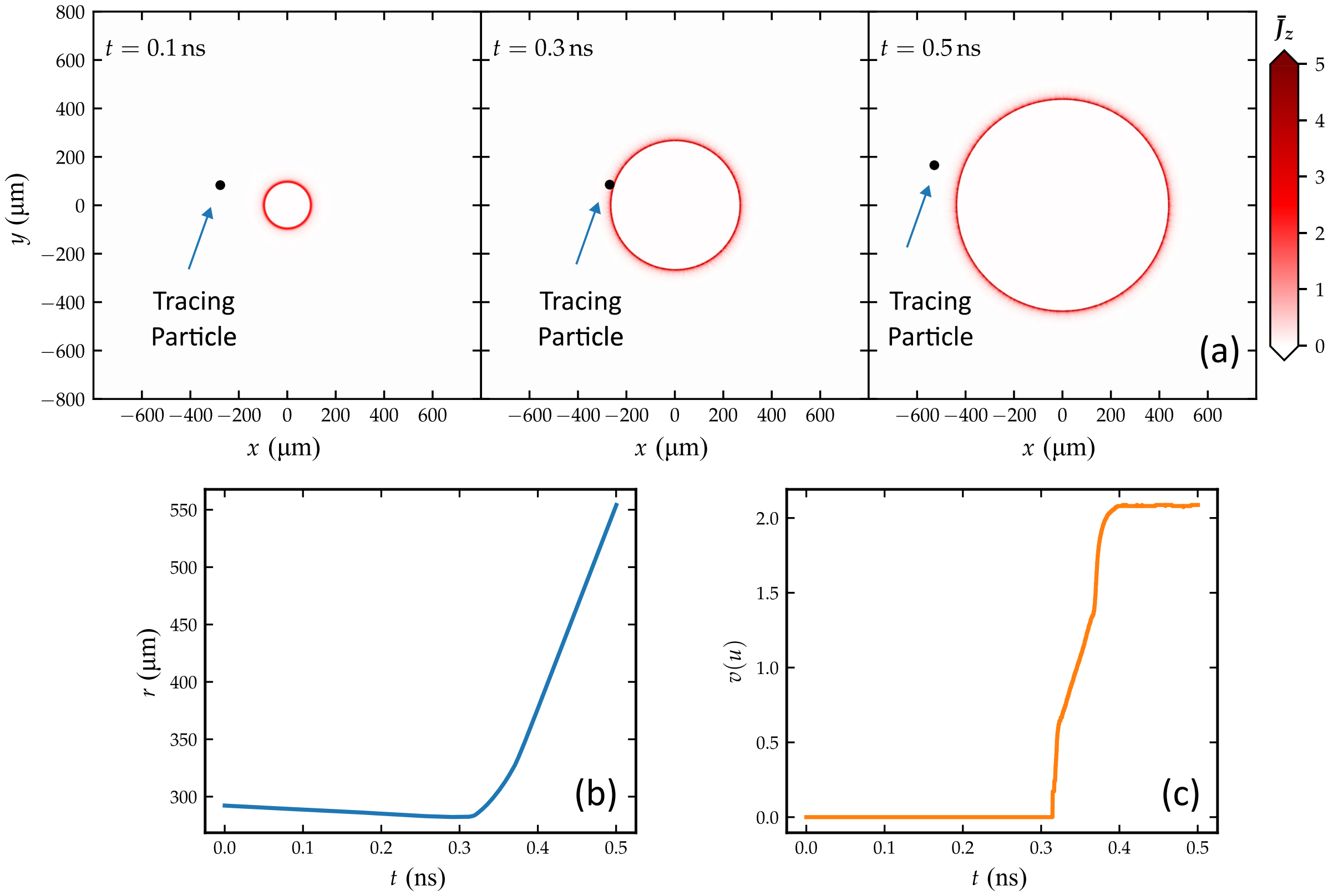

FIG. 5. Particle tracing illustrating ion acceleration through interaction with the expanding front. (a) Trajectory of a representative traced particle together with the position of the moving front at $t = 0.1$, 0.3, and 0.5 ns; the full temporal evolution is provided in the video in the Supplemental Material [58]. (b) Radial position of the particle as a function of time. (c) Particle velocity normalized to the expansion velocity $u$. After interacting with the moving front, the particle is accelerated to approximately $2u$. The acceleration is mediated by the radially outward electric field within the return-current layer. In the frame moving with velocity $u$, the particle velocity changes from $-u$ to $+u$ during the interaction, corresponding to a final velocity of $2u$ in the laboratory frame.

In the thin-layer and large-radius limit ($\delta \ll R$, $a \ll R$), this reduces to

$$\frac{d}{dt}(\rho_0 \pi R^2 U) = \frac{\mu_0 I^2}{4\pi R}. \quad \text{(A6)}$$

Assuming a solution $R = Ut$ and substituting the imposed current profile $I(t) = I_0 t/\tau$, we obtain the scaling for the front velocity:

$$U = \left(\frac{\mu_0 I_0^2}{8\pi^2 \rho_0 \tau^2}\right)^{1/4}. \quad \text{(A7)}$$

In comparison with the simulation, the analytical model reproduces the same scaling dependence on plasma density and current drive. The primary difference lies in the numerical coefficient: 0.28 from the simulation versus 0.33 from the theoretical model. This discrepancy arises from the simplifying assumption of uniform current and plasma distributions. As seen in Fig. 1(b), the current density is not uniform. By numerically integrating the radial magnetic force from the simulation data, we obtain $0.033\mu_0 I^2/R$, whereas the right-hand side of Eq. (A5) gives $0.080\mu_0 I^2/R$. Substituting the numerically integrated magnetic force into Eq. (A5) yields a scaling coefficient of 0.27, in excellent agreement with the simulation result of 0.28.

## APPENDIX B: DETAILS OF THE REVERSE CURRENT AND MAGNETIC-FIELD STRUCTURE AND ION ACCELERATION

This appendix provides supporting simulation results for the reverse-current structure and ion acceleration process discussed in the main text. Figure 4 presents the current cancellation, magnetic-field structure, and physical origin of the reverse current associated with the expanding current layer. Figure 5 illustrates representative ion trajectories through the expanding front and clarifies the origin of the reflected ion population.